\documentclass[12pt]{iopart}
\usepackage[backend=biber,style=numeric-comp, sorting=none, url=false, isbn=false, eprint=false, giveninits=true, doi=true,maxbibnames=50,maxcitenames=1]{biblatex} %Imports biblatex package
\usepackage{graphicx}
\usepackage{todonotes}
\usepackage[separate-uncertainty = true,multi-part-units=single]{siunitx}
\usepackage{multirow}
\DeclareSIUnit{\sample}{S}
\DeclareSIUnit{\slm}{slm}
\DeclareSIUnit{\ppm}{ppm}
\usepackage{booktabs}

\expandafter\let\csname equation*\endcsname\relax
\expandafter\let\csname endequation*\endcsname\relax
\usepackage[version=4]{mhchem} %package for chemical reactions and expressions

\makeatletter
\newcommand\reaction@[1]{\begin{equation}\renewcommand{\theequation}{R\arabic{equation}}\ce{#1}\end{equation}}
\newcommand\reaction@nonumber[1]%
{\begin{equation*}\ce{#1}\end{equation*}}
\newcommand\reaction{\@ifstar{\reaction@nonumber}{\reaction@}}
\makeatother
\newcommand\ExB{$\vec{E}\times\vec{B}$~}

\usepackage{subcaption}

\begin{document}

\title{Azimuthal ion movement in HiPIMS plasmas - Part II: lateral growth fluxes}

\author{Steffen Schüttler\footnote{current affiliation: Plasma Interface Physics, Ruhr University Bochum, Bochum, Germany}, Sascha Thiemann-Monje, Julian Held\footnote{current affiliation: University of Minnesota, Minneapolis, USA}, Achim von Keudell}

\address{Experimental Physics II, Ruhr University Bochum, Bochum, Germany}
\ead{Achim.vonKeudell@rub.de}
\vspace{10pt}
%\begin{indented}
%\item[]January 2023
%\end{indented}

\begin{abstract}
The transport of sputtered species from the target of a magnetron plasma to a collecting surface at the circumference of the plasma is analyzed using a particle tracer technique. A small chromium insert at the racetrack position inside a titanium target is used as the source of tracer particles, which are redeposited on the collecting surface. The azimuthal velocity of the ions along the racetrack above the target is determined from the Doppler shift of the optical emission lines of titanium and chromium.  The trajectories are reconstructed from an analysis of the transport physics leading to the measured deposition profiles. It is shown that a simple direct-line-of sight re-deposition model can explain the data for low power plasmas (DCMS) and for pulsed high power impulse magnetron plasmas (HiPIMS) by using the Thompson velocity distribution from the sputter process as starting condition. In the case of a HiPIMS plasma, the drag force exerted on the ions and neutrals by the electron Hall current has to be included causing an azimuthal displacement in \ExB direction. Nevertheless, the Thompson sputter distribution remains preserved for 50\% of the re-deposited growth flux. The implications for the understanding of transport processes in magnetron plasmas are discussed.
\end{abstract}

%\maketitle
% \ioptwocol
%
% \listoftodos

\section{Introduction}

Magnetron sputtering plasmas are frequently used for thin film deposition for a wide range of applications ranging from hard coatings over thermal insulation layers to microelectronics \cite{gudmundsson_foundations_2022}. In a magnetron sputtering plasma, electrons are confined by a magnetic field in an \ExB configuration. The unmagnetized ions are attracted by the electric field towards the target acting as the cathode. The ions sputter the target material, forming the so called racetrack underneath the plasma torus. Sputtered particles move through the discharge to the substrate, forming a film there. In a conventional direct current magnetron sputtering (DCMS) plasma operated at low power densities at the target, the film forming particles are mostly neutral atoms of the target material. In the case of high power impulse magnetron sputtering (HiPIMS), much higher voltages are used to drive large currents, leading to ionization of the sputtered material. Coatings are then created by energetic ions instead of neutrals, leading to superior coatings \cite{daniel_adhesion_2020, alami_ion-assisted_2005,samuelsson_film_2010}. However, the discharge is then operated in short high voltage pulses with lengths on the order of \SI{100}{\micro\second} and  duty cycles in the order of 1\% to prevent melting of the target material. The major drawback of HiPIMS is that it often leads to reduced deposition rates compared to DCMS for the same average power, since most of the sputtered species are ionized and return to the target, as they are susceptible to the target-directed electric fields in the magnetic trap region \cite{brenning_understanding_2012, anders_deposition_2010,samuelsson_film_2010}. Consequently, understanding the transport from target to substrate is of key importance for the optimization of HiPIMS plasmas. 

In HiPIMS plasmas, these ion transport processes are very complex and difficult to study. The most important aspect that has traditionally received the most attention is the strong electric field present in the magnetic trap region of the discharge, pulling ions back towards the target surface. This electric field has been measured to be a few \SI{1000}{\volt\per\meter} \cite{rauch_plasma_2012, mishra_evolution_2010} and is oriented perpendicular to the magnetic field lines, pointing towards the racetrack \cite{rauch_plasma_2012, mishra_2d_2011}. Ions are also affected by collisions \cite{breilmann_fast_2017,desecures_determination_2017,trieschmann_neutral_2017}, mostly ion-ion Coulomb collisions in the case of high ion density \cite{held_velocity_2020,breilmann_influence_2017}. Finally, the ion transport is also affected by the presence of spokes, long-wavelength oscillations observed in magnetron sputtering plasmas \cite{hecimovic_spokes_2018, held_pattern_2020}. Spokes have been shown to modulate the electric field strength and direction \cite{panjan_plasma_2017, lockwood_estrin_triple_2017, held_electron_2020, held_spoke-resolved_2022, boeuf_spoke_2023} which can affect the trajectory of ions \cite{anders_drifting_2013,panjan_asymmetric_2014,breilmann_fast_2017}.

The transport of species from target to substrate is often analyzed by passive and active optical spectroscopy to evaluate distribution functions and densities at various locations in the discharge \cite{britun_optical_2020, britun_time-resolved_2012, held_velocity_2020, hnilica_revisiting_2020, hnilica_revisiting_2020-1, palmucci_rarefaction_2013, kanitz_two_2016, el_farsy_characterization_2019, desecures_characterization_2014}. Alternatively, ion mass spectroscopy is used to monitor the growth flux at the substrate position \cite{palmucci_mass_2013, hippler_pressure_2019,breilmann_dynamic_2013}. However, determining transport properties from measurements at the substrate position can be challenging, since, for example, a specific mass spectrometry signal measured at the substrate position at a particular time may be composed of species starting at different target positions  follow different trajectories and transit times. This challenge also occurs for any optical diagnostics with spatial and time resolution like laser induced fluorescence, where each probed volume element within the discharge contains particles with different origins, rendering the interpretation of the measurements highly challenging \cite{el_farsy_characterization_2019,hnilica_revisiting_2020-1}. Therefore, we here use a \textit{particle tracer technique} to evaluate species transport in complex plasmas and/or complex geometries where particles start from a well defined local position given by an insert in the target to a collector position. Such approach is often being used in nuclear fusion research to follow re-deposition patterns of locally released particles \cite{kirschner_modelling_2004}.

In our implementation of the particle tracer technique, we use an insert of chromium at a specific location on the racetrack of a titanium target and collecting film deposition at samples around the circumference of the magnetron on a collector. The simplest transport model would be a direct-line-of sight deposition from the insert onto the collecting surface at the circumference, which only accounts for the cosine distribution of the sputter process, the absolute distance between sputter location and collecting location and the cosine projection of the adsorption process. Any deviation from this, can be used to deduce important forces acting on the species on their path from target to collecting location. Most prominent is the friction force from the Hall current along the \ExB direction that induces an azimuthal velocity component of the sputtered species. By following the particle trajectories the impact of a HiPIMS plasma on the growth flux is analysed and discussed.  

This azimuthal movement of ions in magnetron plasmas is addressed in a two part series with part I addressing the velocity distribution functions of the ions inside the plasma\cite{thiemann-monje_azimuthal_2023}. This is part II focusing on the lateral deposition of species leaving the magnetic trap region. 

\section{Experiments} 

\subsection{Plasma source}

The 2" magnetron source (Thin Film Consulting IX2U) was radially symmetric and suitable for round targets with a radius of $r_{target}=\SI{25}{\mm}$, resulting in a target area of about \SI{20}{\cm^2}. Titanium (purity of \SI{99.99}{\percent}) was used as the bulk target material. The center of the racetrack on the target was located at a radius of $r_{racetrack}=\SI{13.5}{\mm}$. At this position, an insert of chromium (purity \SI{99.95}{\percent}) with a diameter of \SI{6}{\mm} was mounted within the target, as previously discussed in \cite{layes_composite_2017,layes_species_2017}. The magnetron was placed inside a vacuum chamber with a base pressure of \SI{4e-6}{\pascal}. The discharges were operated in argon (purity of \SI{99.999}{\percent}) at a pressure of \SI{0.5}{\Pa}. The working gas was injected via a gas shower located at the top of the vacuum chamber. A thin metallic ring was used as the anode cover, to allow optical access to the plasma emission originating close to the target surface, as well as an almost unobstructed ion transport in radial direction. The position of the anode cover and the magnetic field configuration of the employed magnetron can be found in a recent publication \cite{held_velocity_2020}. 
%A grounded anode cover was placed on the side above the target with adjustable distance to the target. A typical distance of 4 mm to the target surface was chosen. The magnetic field was reconstructed from Hall probe measurements using the approach by \citeauthor{kruger_reconstruction_2018} \cite{kruger_reconstruction_2018}. The magnetic field strength in the vicinity of the target reaches values of about \SI{0.1}{\tesla}.

A cylindrical coordinate system is used with the z-coordinate pointing away from the target in normal direction, while the r-coordinate describes the radial distance to the center of the magnetron. The azimuthal direction $\Phi$ points in \ExB electron drift direction. We define the position of the chromium insert as $\Phi = \SI{0}{\degree}$.

Experiments were performed for DCMS as well as for HiPIMS plasmas. In case of the HiPIMS discharge, the power supply (Trumpf TruPlasma Highpulse 4002) was connected via an external inductance to the magnetron, which limits the current rise during discharge ignition preventing discharge runaway at high powers \cite{breilmann_influence_2017}. Discharge current (Tektronix TCP 404) and voltage (Tektronix P6015A) were measured in the cable between the external inductance and the magnetron. For DCMS, another power supply (Trumpf TruPlasma Highpulse 4001 G2) was used without external inductance. The voltage measurement was performed in the same way as for the HiPIMS discharge, whereas the much smaller current was measured using a more sensitive current probe (ELDITEST CP6550). The electrical measurements were taken using an oscilloscope (Tektronix TPS 2024) triggered on the beginning of the voltage pulse. 

The HiPIMS discharge was pulsed with a frequency of \SI{40}{\Hz} and a pulse width of \SI{100}{\micro\s}. The applied voltage was \SI{500}{\V} after breakdown, which occurred at a delay of about \SI{20}{\micro\s}. The current increased after breakdown and reached a maximum value of \SI{50}{\A} at the end of the pulse. The peak power in the HiPIMS discharge was \SI{25}{\kW} with a peak power density of \SI{1.25}{\kW\per\cm^2} and an averaged power density of \SI{2.4}{\W\per\cm^2}. Within the accuracy of the electrical measurements, the current and voltage time traces are identical to the ones obtained without the Cr insert, published in a previous article \cite{held_velocity_2018} (\SI{590}{\volt} case).  The DCMS discharge was operated at \SI{290}{\V} and \SI{65}{\mA} resulting in a power of \SI{19}{\W} and a power density of about \SI{1}{\W\per\cm^2}. In the following, we only refer to the \textit{HiPIMS plasma} and \textit{DCMS plasma} at these specific operating parameters.

\subsection{Particle tracking procedure}

% \begin{figure}[htb]
%     \centering
%     \includegraphics[width=0.7\linewidth]{figures/Sketch_Coating_crop.pdf}
%     \caption{Sketch of the plasma chamber with a circular collector around the magnetron housing the wafers where the redeposition is monitored.}
%     \label{fig_setup:coating}
% \end{figure}

\begin{figure}[h]
    \centering
    \begin{subfigure}[][][t]{0.59\textwidth}
         \centering
         \includegraphics[width=\textwidth]{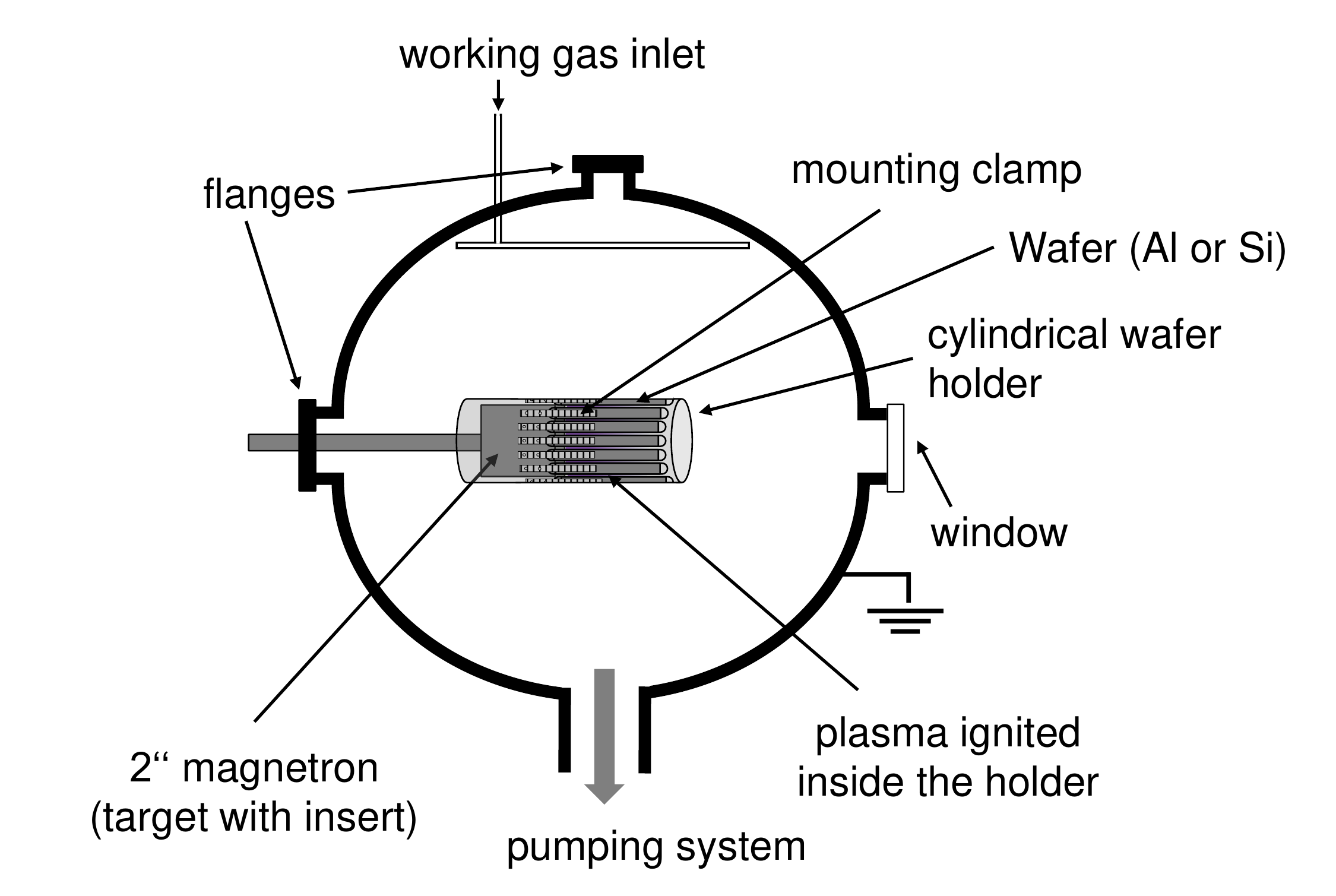}
    \end{subfigure}
    \begin{subfigure}[][][t]{0.4\textwidth}
         \centering
         \includegraphics[width=0.8\textwidth]{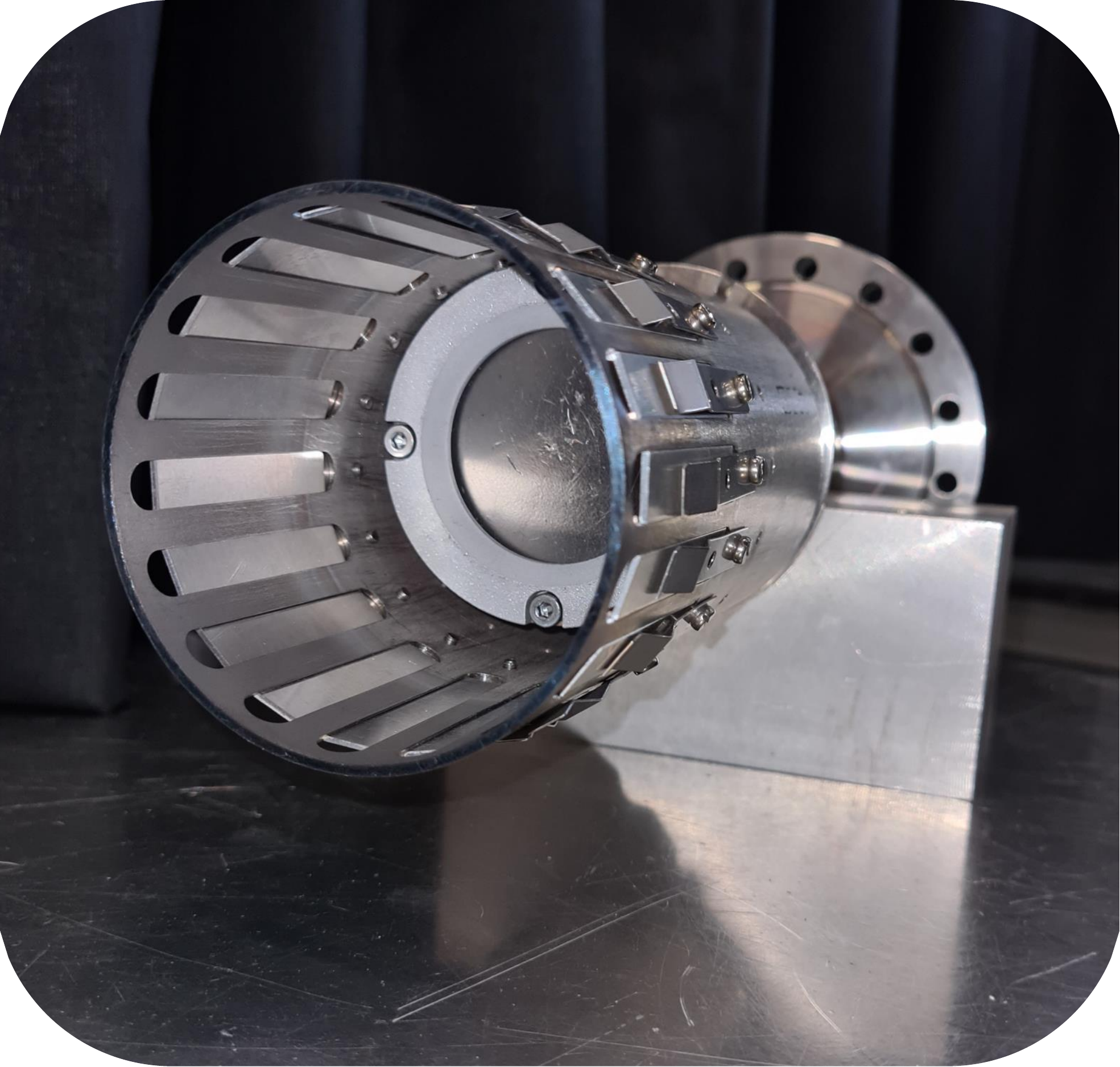}
    \end{subfigure}
    \caption{(left) Sketch of the plasma chamber with a circular collector around the magnetron housing the wafers where the re-deposition is monitored. (right) Picture of the collector mounted to the magnetron.}
    \label{fig_setup:coating}
\end{figure}

\begin{figure}[htb]
    \centering
    \includegraphics[width=0.4\textwidth]{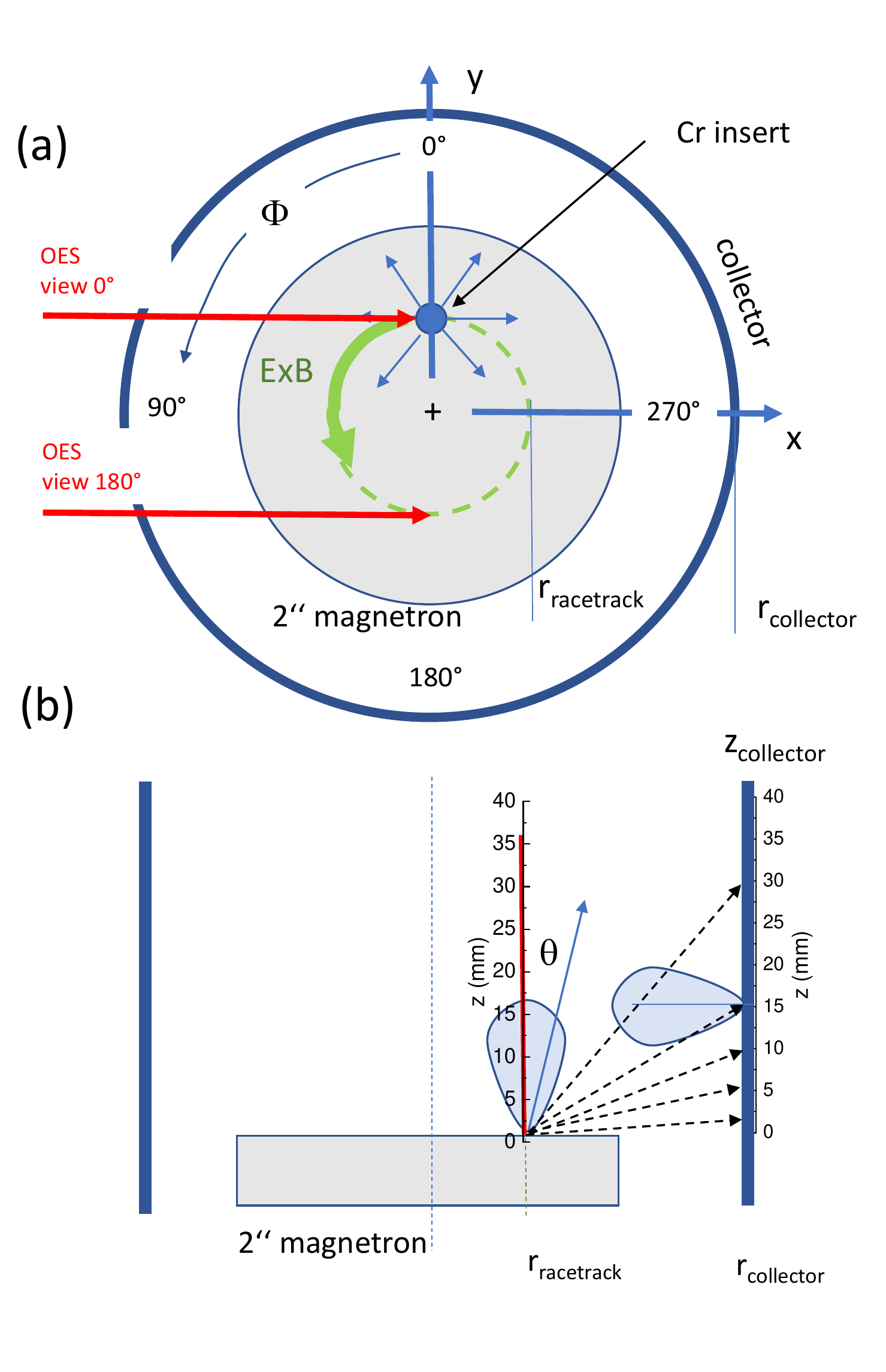}
    \caption{Particle tracing concept. (a) Top view of the 2'' magnetron with the location of the racetrack and the collector. The \ExB direction is indicated. The red arrows denote the line-of-sight of the OES measurement at 0° and at 180°. (b) Side view of the 2'' magnetron. The line-of-sight trajectories of the sputtered species leading to deposition on the collector are shown as dashed lines. The dimensions are to scale.}  \label{fig_setup:Particle_tracing_concept}
\end{figure}

To track the movement of sputtered particles, the material deposition on wafers mounted on a collector around the magnetron was analyzed. The magnetron was inserted into the vacuum chamber from the side, as shown in figure \ref{fig_setup:coating}. A cylindrical collector made of stainless steel was placed around the magnetron. The collector was \SI{80}{\milli\meter} in diameter or with a radius of $r_{collector}$ = \SI{40}{\milli\meter} and \SI{150}{\milli\meter} in height, allowing us to be able to also examine any coatings at large distances from the target. A total of 18 aluminium or silicon wafers (\SI{10}{\mm} x \SI{50}{\mm}) were mounted on the inside of the collector, every \SI{20}{\degree}. The collector setup is also shown schematically in Fig. \ref{fig_setup:Particle_tracing_concept} with the coordinate system illustrated in the $xy$-plane (a) and the $yz$-plane (b). The viewing direction of the optical emission spectroscopy are indicated as red lines in (a), the \ExB direction is marked. The different possible direct line-of-sight trajectories in the $yz$-plane are illustrated by dashed lines in (b) including the cosine (light blue) distribution for particle sputtering at the target and 
adsorption at the collector. The sketches in Fig. \ref{fig_setup:Particle_tracing_concept} are to scale.   

Depositions on aluminum were used to create samples for ex-situ XPS analysis to avoid any charging of the samples during acquisition of the XPS spectra. Silicon has been used as substrate to measure the coating thickness by profilometry. Comparing current-voltage measurements with and without collector, we found the presence of the grounded collector to not strongly affect the plasma. 
 
\subsection{Surface analysis}

A profilometer (Dektak 6M, Digital Instruments Veeco Metrology Group) was used to measure the film thickness on the coated wafers. During the coating process, one half of the silicon wafers was covered with Kapton tape to obtain a sharp edge of the coating on the silicon wafers. A scale on the wafer indicated the height of the measurement position on the wafer to obtain the coating thickness as a function of the distance $z$ to the target surface. The stylus force was kept to \SI{10}{\milli \g} while the measurement range was set to \SI{65}{\kilo \angstrom} in the "Hills \& Valleys" profile mode. The measurement length was \SI{250}{\micro \m} by a measurement time of \SI{30}{\s}. The coating procedure and measurement of the film thickness were repeated three times and the mean and standard deviation were determined.

The composition of the deposited material on the wafers was analysed by XPS (PHI Versaprobe 5000 spectrometer). The 2p photoelectron peaks of chromium and titanium were investigated with a pass energy of \SI{46.95}{eV} and a resolution of \SI{0.075}{eV}. XPS spectra were taken at five different positions on the wafers, corresponding to five different distances to the target surface. The background of the spectra was fitted by the Shirley method and the relative concentrations of the species were determined. Since chromium was used only as insert material, the concentration of chromium on the wafers was low. Thus, the signal-to-noise ration of the spectra was very low and the background fitting was repeated ten times to reach a standard deviation of \SI{8}{\percent}. The 2p peak of aluminium was also recorded for reference at each measurement position. No aluminium was detectable, indicating a complete deposited layer of titanium and chromium on the wafers. 

\subsection{Rotational Ion velocity}

Measurements of the rotational ion velocity by optical emission spectroscopy were performed, as described previously \cite{held_velocity_2018,held_velocity_2020}. The light emission of the plasma was coupled into an optical fiber (diameter of \SI{880}{\micro m}) using a lens (focal length of \SI{150}{mm}). The lens limited the field of view of the optical setup to a cone of approximately \SI{2}{mm} diameter inside the plasma. Both fiber and lens were mounted on a stage to be able to move the field of view parallel ($y$ direction) and perpendicular (z direction) to the target surface. 

The emission lines of titanium ions and chromium ions were recorded with an intensified CCD-Camera (Andor iStar DH320T-25U-A3) attached to a plane grating spectrograph (Zeiss PGS 2, \SI{1300}{lines/mm} grating, \SI{2}{m}). A spectral resolution of \SI{1.5}{pm} pixel-to-pixel on the camera chip was achieved by operating the spectrometer in the third diffraction order. A gate delay of \SI{85}{\micro s} was set to collect the light of the last \SI{15}{\micro s} of the HiPIMS pulse. Besides the plasma emission, the emission of a hollow cathode lamp (HCL, Cathodeon 3UNX Cr) was measured simultaneously to provide an unshifted wavelength reference to calibrate the plasma emission.

In table \ref{tab:OES_lines}, the investigated ion and neutral lines of chromium and titanium are listed. Since the spectrometer covered a range of about \SI{1.5}{\nm} all lines could be monitored at the same time. The chromium neutral line  was used to determine the instrumental profile of the spectrometer and the reference wavelength from the HCL emission.

\begin{table}[htb]
    \centering
    \setlength{\tabcolsep}{10pt}
    \renewcommand{\arraystretch}{1.2}
    \caption{Data on the optical transitions used for the OES measurements \cite{saloman_energy_2012,saloman_energy_2012-1}.}
    \hspace{3mm}
    \begin{tabular}{rlccc}
    \toprule
    & Species & Ti II & Cr II & Cr I\\
    & Wavelength [nm] & 454.962   & 455.864 & 454.595 \\
    \midrule
    \parbox[t]{3mm}{\multirow{3}{*}{\rotatebox[origin=c]{90}{upper }}} &
   % \multicolumn{4}{c}{\underline{\textbf{Upper level}}} \\
     Energy [eV] & 4.31 & 6.97 & 3.67 \\
    & Configuration & $3d^3(^4F)4p$ & $3d^4(^5D)4p$ & $3d^4(^5D)4s4p(^3P^{\circ})$ \\
    & J & 9/2 & 7/2 & 2 \\
    \midrule
    \parbox[t]{4mm}{\multirow{3}{*}{\rotatebox[origin=c]{90}{lower }}} & Energy [eV] & 1.58 & 4.07 & 0.94\\
    & Configuration & $3d^3$ & $3d^5$ & $3d^5(^6S)4s$\\
    & J & 11/2 & 9/2 & 2 \\
    \bottomrule 
    \end{tabular}
    \label{tab:OES_lines}
\end{table}

The determination of the velocity distribution function (VDF) from the emission lines was performed, as described in \cite{held_velocity_2018}. The emission lines in HiPIMS discharges are mainly broadened by Doppler broadening and instrumental broadening. To remove the latter, a Wiener deconvolution was used. The wavelength axis of the spectra can be transformed into a velocity axis by $v = c(\lambda/\lambda_0 - 1)$, with $\lambda_0$ being the unshifted wavelength of the emission line deduced from the HCL measurement. The average rotational velocity of the ions is then given by the mean velocity of the shifted VDFs. At each measurement position, the emission spectrum was recorded three times and the mean value and standard deviation were calculated.

\section{Results and Discussion} 

\subsection{Rotational ion velocities}

\begin{figure}
    \centering  \includegraphics[width=0.5\linewidth]{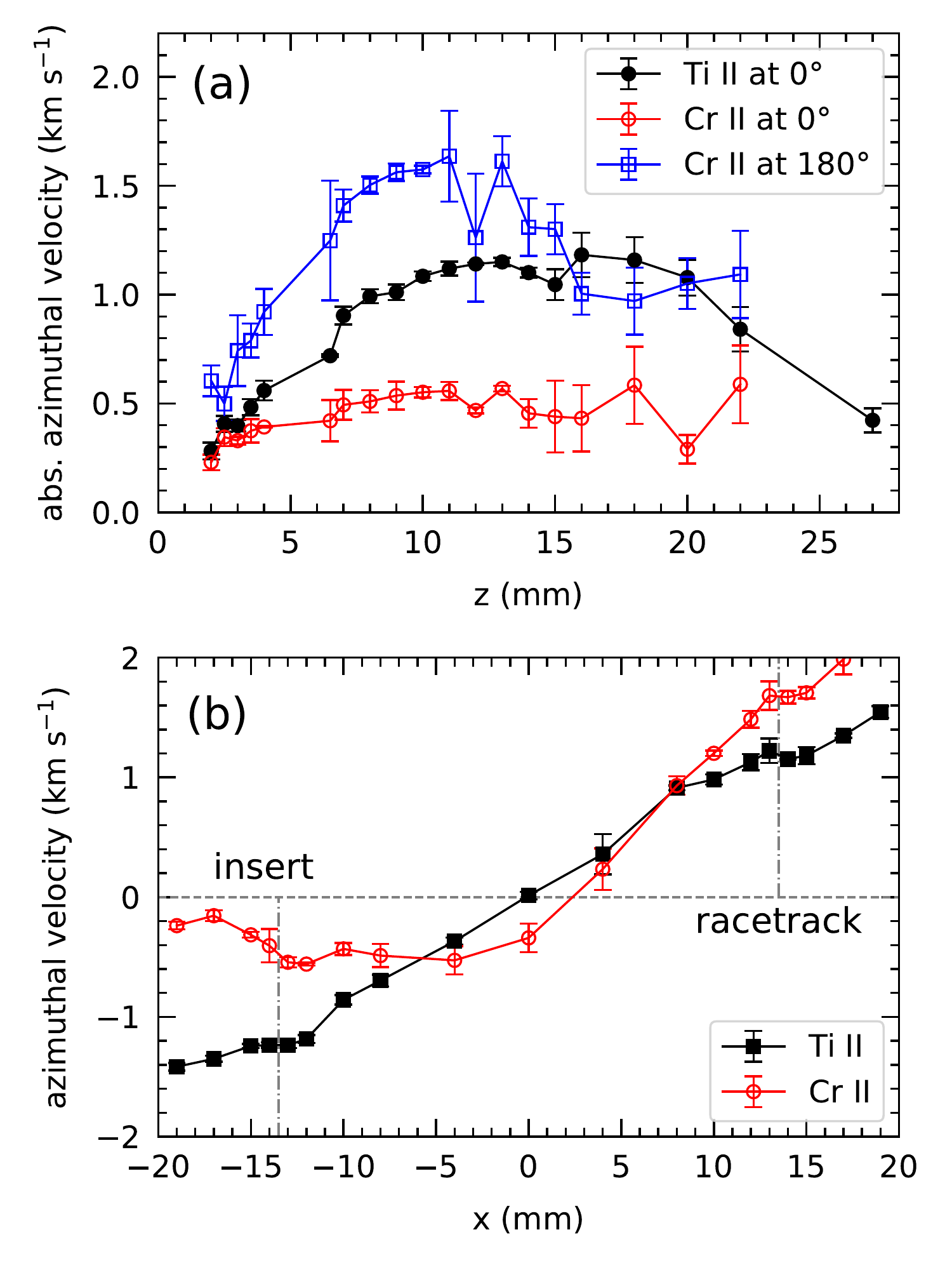}
    \caption{Average azimuthal velocities of Ti II and Cr II at (a) different distances $z$ to the target above the racetrack measured at 0° and at 180° and $z=$ 13.5 mm and (b) across the target at $z=$ 15 mm in the HiPIMS plasma. The location of the racetrack is denoted as vertical dashed lines and the position of the insert is indicated in (b).}
    \label{fig:vrot}
\end{figure}

The high resolution optical spectra were used to determine the velocity distribution function (VDF) of titanium and chromium ions in the HiPIMS plasma. From the VDF, the average azimuthal ion velocity was calculated for both titanium as well as chromium ions at different positions inside the plasma. Fig. \ref{fig:vrot}a shows the variation of the average rotational velocities with distance $z$ to the target measured at 0° (insert position) and at 180° (opposite insert) above the racetrack for both titanium (Ti\,II) as well as chromium ions (Cr\,II). For both species and positions, the average velocity initially increases with target distance, showing a maximum between $z=\SI{10}{\milli\meter}$ to $z=\SI{15}{\milli\meter}$ and then decreases again. %The shape and the values of the titanium ion azimuthal velocity fit well to the measurements shown in part I of this paper series, indicating that the titanium discharge might not critically altered by the presence of the small amount of chromium. 
For the chromium ion movement measured at 0° and at 180°, an asymmetry is visible on the two sides of the target, with the velocities at 0°, directly above the insert, being much smaller than the velocities at 180° on the opposite side. This is easy to understand since the chromium atoms, after being sputtered require some path length to become ionized and accelerated to their final speed.

% \begin{figure}
%     \centering  \includegraphics[width=0.5\linewidth]{figures/Rotational_velocities.pdf}
%     \caption{Average rotational velocities of Ti II (a) and Cr II (b) at different distances $z$ to the target above the racetrack measured at 0° and at 180°.}
%     \label{fig:vrot_z}
% \end{figure}

% \begin{figure}
%     \centering
%     \includegraphics[width=0.5\linewidth]{figures/Rotational_velocities_radial.pdf}
%     \caption{Scan of the average rotational velocity across the target for titanium (a) and chromium from the insert (b). The location of the racetrack is denoted as vertical dashed lines. Th position of the insert is indicated.}
%     \label{fig:vrot_r}
% \end{figure}

Fig. \ref{fig:vrot}b shows the average rotational velocities of both metal ions being scanned parallel to the target surface at a height of $z=$ 15 mm. The variation of the average rotational velocity for titanium shows positive and negative velocities on opposite sides of the target center. There, the velocities at 0° and at 180° are exactly the opposite, because the line-of-sight of the OES measurement is observing either against or with the \ExB movement of the ions, respectively. The velocity is continuously increasing with distance to the target center, which indicates that the angular velocity $\omega_{Hall}$ of the Hall current is almost independent of the radius and accelerates the ions at larger distances to the target center much faster. The Hall current is expected to follow the \ExB drift of the electrons with a magnitude scaling with $E/B$. Since the $E$ field is constant at a particular distance to the target surface and only the $B$ field decreases almost linearly with distance to the target center, the \ExB drift is expected to also increase linearly with distance to the target center. As a result, the ion velocity follows the \ExB drift of the electrons and can be simply expressed as $v_{ions} \propto \omega_{Hall} \times r$ with a radius independent $\omega_{Hall}$. The variation of the average rotational velocity for chromium parallel to the target surface is different. Above the insert the velocity is only \SI{-0.5}{km\per s} and increases to \SI{+1.3}{km\per s} at the opposite side above the racetrack. Apparently, the sputtered chromium species from the insert are accelerated by \SI{0.8}{km\per s} when passing 180° along the racetrack.

\begin{figure}
    \centering  \includegraphics[width=0.5\linewidth]{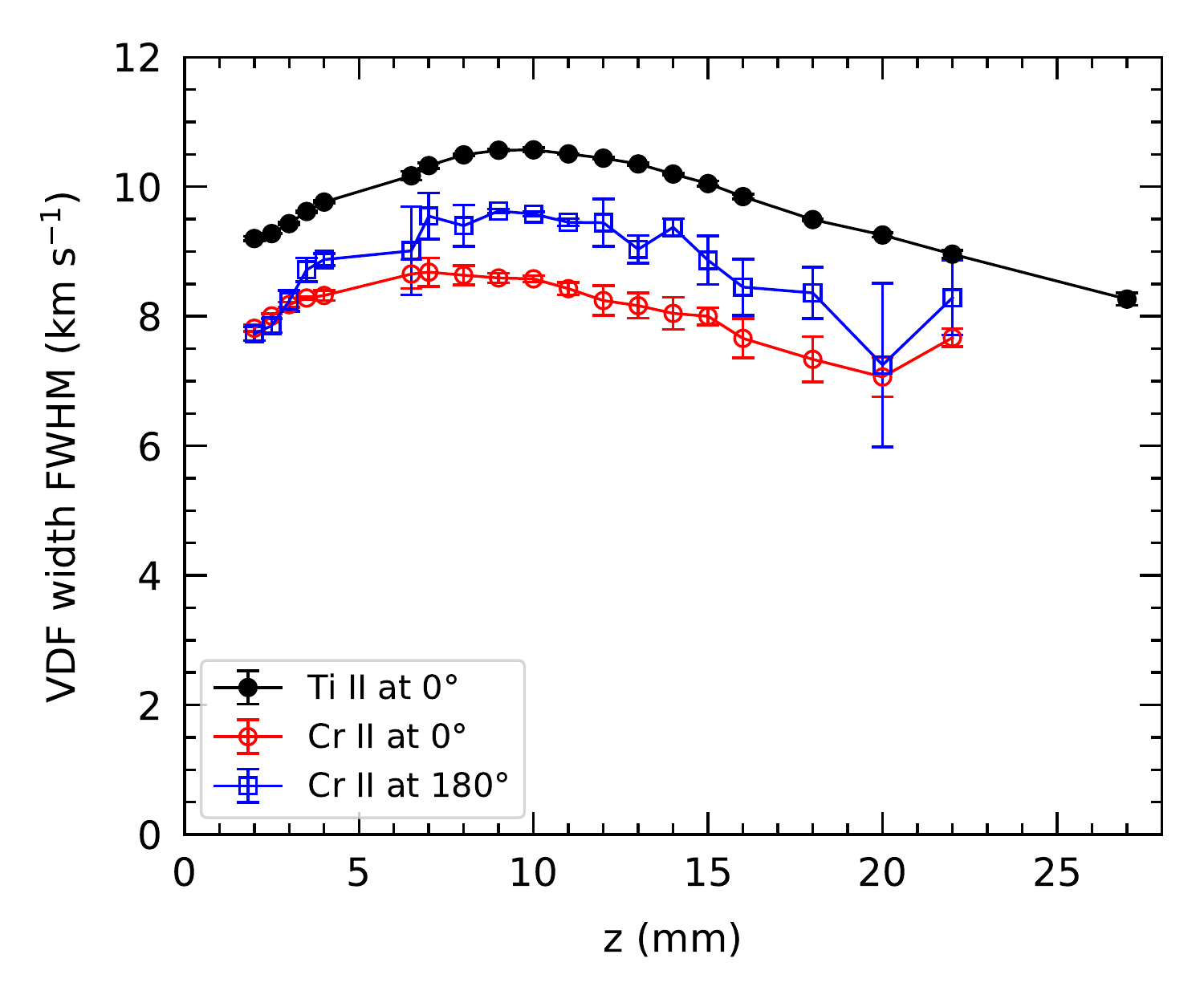}
    \caption{Width of the VDFs for chromium and titanium ions at different heights $z$ above the target surface in the HiPIMS plasma.}
    \label{fig:VDFs}
\end{figure}

It is interesting to note that the velocities of chromium ions are a bit larger than for titanium ions. Apparently, the acceleration by the electron Hall current is more efficient for chromium. This could be explained by the higher mass of chromium compared to titanium. Thereby, the chromium ions preserve more easily their forward momentum being transferred from the electrons. This hypothesis is evaluated by regarding the width of the VDFs of chromium above the insert, and at 180° as well as the width of the VDF of titanium as shown in Fig. \ref{fig:VDFs}. It can be seen that the width of the VDF is the smallest for chromium ions above the insert at 0° and becomes broader to 180° after the sputtered chromium species are ionized and accelerated. The width of the chromium ion VDF remains smaller than the width of the titanium ion VDF. This is consistent with our arguments above. The variation of the width of the VDF with distance $z$ is similar to the variation of the average rotational velocity with distance $z$, showing the largest width at $z=$ 15 mm.   

\subsection{Radial deposition}

\begin{figure}
    \centering
    \includegraphics[width=0.5\linewidth]{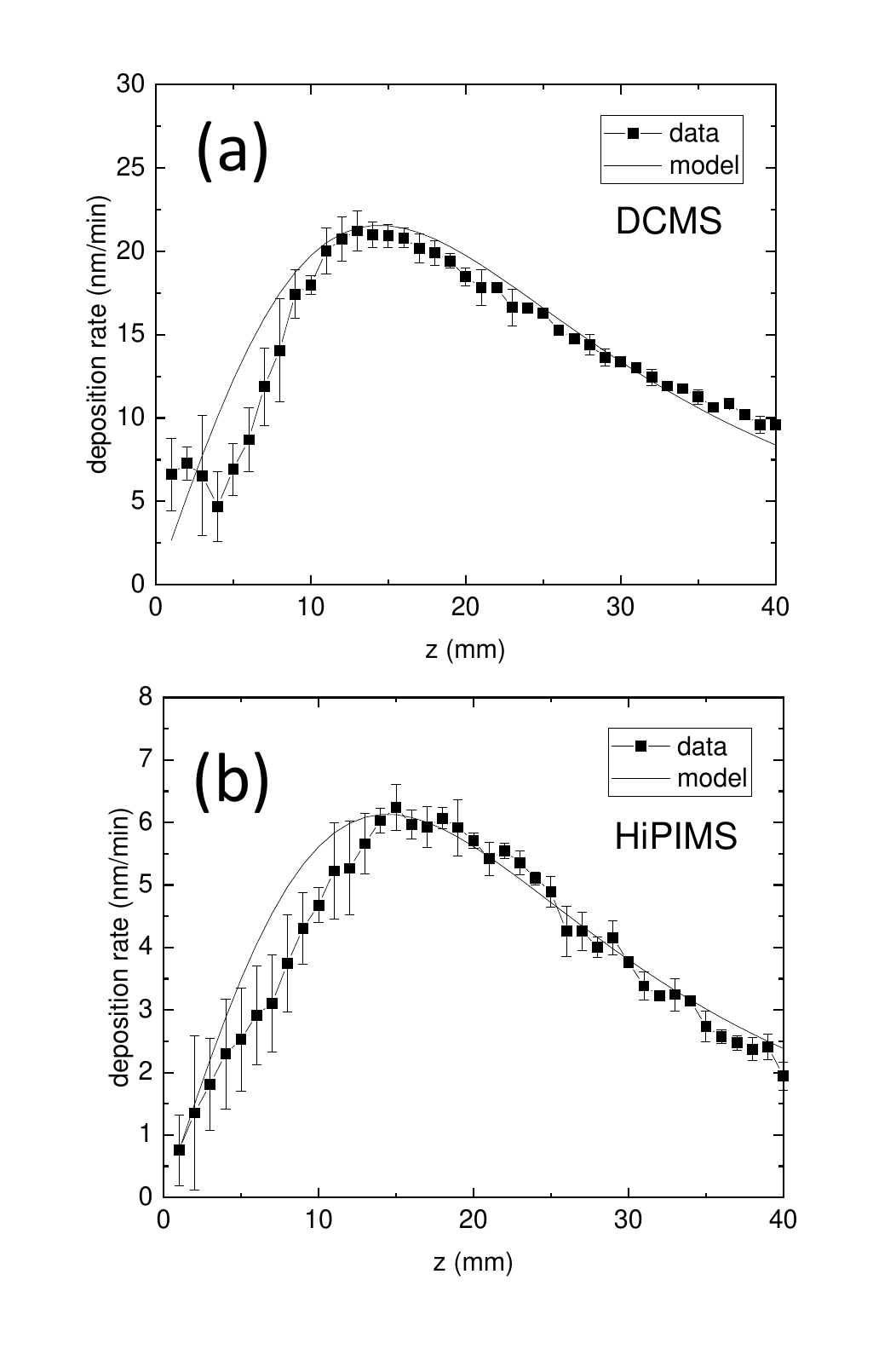}
    \caption{Growth rates measured at different heights $z$ above the target surface on the collector for a DCMS (a) and a HiPIMS (a) plasma. The solid line refers to a simple line-of-sight deposition model.}
    \label{fig:sideways_deposition_model}
\end{figure}

At first we regard the overall radial deposition rate on the collector. Since the surface area of the titanium bulk target is much larger than the small insert made of chromium, the total deposition rate is dominated by titanium. We therefore find no variations of the total deposition rate in azimuthal direction $\Phi$ and only consider the  variation in the target normal direction $z$. For both HiPIMS and DCMS, the radial deposition rate measured on the collector wafer at different heights $z$ above the target surface is shown in Fig. \ref{fig:sideways_deposition_model}. For the HiPIMS discharge, the radial deposition rate increases with target distance, peaks at around \SI{15}{\milli\meter} at \SI{6}{\nano\meter\per\minute} and then decreases. For DCMS, a dip in deposition rate around $z = \SI{5}{\milli\meter}$ can be observed, caused by the anode cover blocking some of the particles from reaching the substrate. Otherwise, the same trend is observed as for the HiPIMS case, but with a generally higher deposition rate peaking at \SI{21}{\nano\meter\per\minute}. These growth rates are in good qualitative agreement with recent results of Hajihoseini \etal obtained under similar condition \cite{hajihoseini_sideways_2020}. 

If we normalize this difference in deposition rate to the average power (in the HiPIMS case 2.4 W/cm$^2$ and in the DCMS 1 W/cm$^2$ ), we obtain a factor of about 8.8 higher deposition rate for DCMS compared to HiPIMS. This is larger than the ratio 3, as being reported by Samuelsson et al. \cite{samuelsson_film_2010}, who analyzed a titanium magnetron plasma for an average power of 2.7 W/cm$^2$, but at a much smaller peak power 0.18 kW/cm$^2$. This difference in peak power can explain the different ratios, because a high peak power implies a higher ionization degree and a more dominant return effect, which renders our HiPIMS growth rates much smaller. 

The shapes of the spatial deposition profiles are compared to a simple collisionless model, projecting the cosine distribution of sputtered species emitted at the racetrack position to the collector position at different heights $z$ leading to different growth rates depending on the angle of incidence. Sputtered particles emitted at an angle $\theta$ are projected to the deposition on the wafer at an angle $\pi/2-\theta$ to the wafer normal (see also Fig. \ref{fig:sideways_deposition_model}b). If the location of emission and the location on the wafer are at a distance $l$ from each other, the growth rate $\Gamma$ scales than as:

\begin{equation}
\Gamma \propto \cos(\theta) \frac{1}{l^2} \cos\left(\frac{\pi}{2}-\theta\right)\label{eq:depo} ~ .
\end{equation}

The angle $\theta$ can be expressed in coordinates $r$ and $z$ and we obtain with $l^2=z^2+(r_{collector}-r_{racetrack})^2$:

\begin{equation}
\Gamma(z) \propto \frac{z(r_{collector}-r_{racetrack})}{(z^2+(r_{collector}-r_{racetrack})^2)^2}
\end{equation}

This simple model is shown for comparison in Fig. \ref{fig:sideways_deposition_model} showing excellent agreement. The very simple direct line-of-sight deposition model can accurately describe the growth rates on the collector. %This line-of-sight modeling neglects any azimuthal movement of the growth species that maybe induced by electric fields in the plasma for the ions or by the friction with the Hall current in the plasma torus. The good agreement between data and model shows that such effects are of second order, but may explain the small deviations that are still visible. In case of an azimuthal movement of the species, the actual travel distance $l$ along the direct line-of-sight in eq. \ref{eq:depo} becomes a bit larger and the growth rate is slightly reduced (the change in the emission and adsorption angles can be neglected). This might explain the deviation at large distances for the HiPIMS plasma, where deposition is dominated by ions, which are most susceptible to any azimuthal forces. 
Deviations at smaller distances $z$ for both cases of the HiPIMS and the DCMS plasma are maybe an artefact due to the anode cover at $z$= 4mm partially blocking the direct line-of-sight. It should be mentioned that the analysis of the azimuthal deposition rates reveal a 50\% un-directed background deposition rate from scattered species, as presented below. Such an un-directed background deposition rate for titanium could became visible as a constant offset in the $z$ variation of the growth rate shown in Fig. \ref{fig:sideways_deposition_model}. However, also an un-directed background from scattered species will follow the same scaling as given by eq. \ref{eq:depo}.

When using the insert in the target, the local measurement of chromium on the collector allows to also analyze the azimuthal transport of species. To this end, the chromium percentage within the deposited material was determined using XPS, allowing us to calculate the chromium deposition rate from the chromium percentage and the total deposition rate deduced from the film thickness.

\begin{figure}
    \centering
    \includegraphics[width=0.5\textwidth]{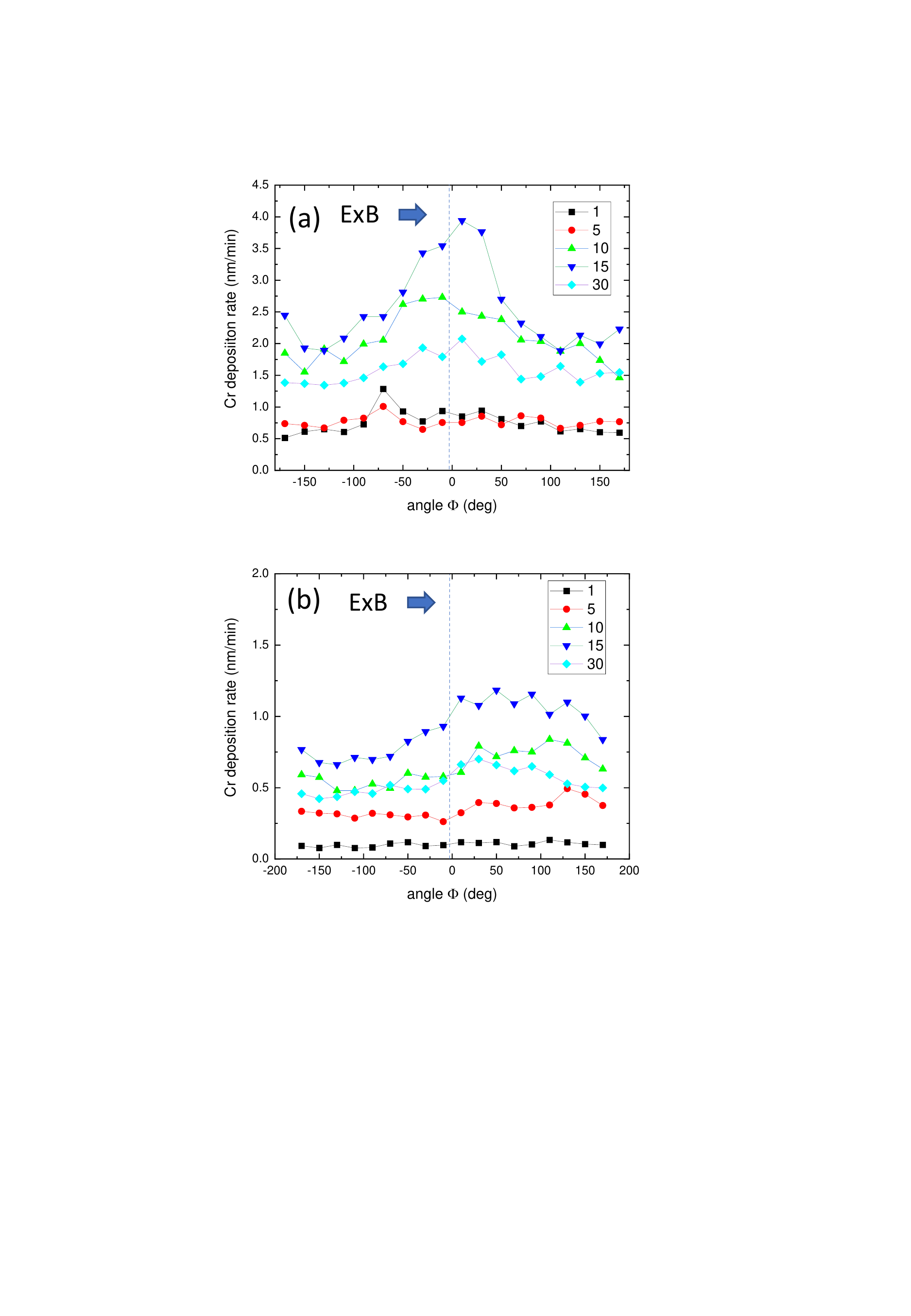}
    \caption{Deposition rate of the insert material along the circumference of the target at different heights $z$ above the target for the DCMS plasma (a) and the HiPIMS plasma (b).}
    \label{fig:Cr_deposition}
\end{figure}

Fig. \ref{fig:Cr_deposition} shows the chromium deposition rate measured at different heights $z$ above the target surface for the DCMS (a) and the HIPIMS (b) case at different angles $\Phi$ along the circumference. In the HiPIMS case one can clearly see that the deposition is shifted in \ExB direction indicating that the deposition by ions is in fact affected by azimuthal forces from the Hall current exerted on the ions during transport from target to collector. Such a shift is absent in the DCMS case, where the deposition from neutrals dominates and the maximum in deposition occurs in close proximity to the insert position. 

The deposition at different heights $z$, as shown in Fig. \ref{fig:sideways_deposition_model}, could be very well explained by a line-of-sight deposition model. Therefore, it is straightforward to also test the azimuthal distribution using a line-of-sight deposition model. Such a simple model neglects any scattering in the discharge or any bending of ion trajectories by electric fields in the plasma. The shift towards the \ExB direction in the HiPIMS case can be accounted for by adding an azimuthal velocity that displaces a typical deposition profile, as can be measured for the DCMS plasma, along the circumference during the transit time of a species from target to collector. This is discussed in the following.

\begin{figure}
    \centering
  \includegraphics[width=0.5\textwidth]{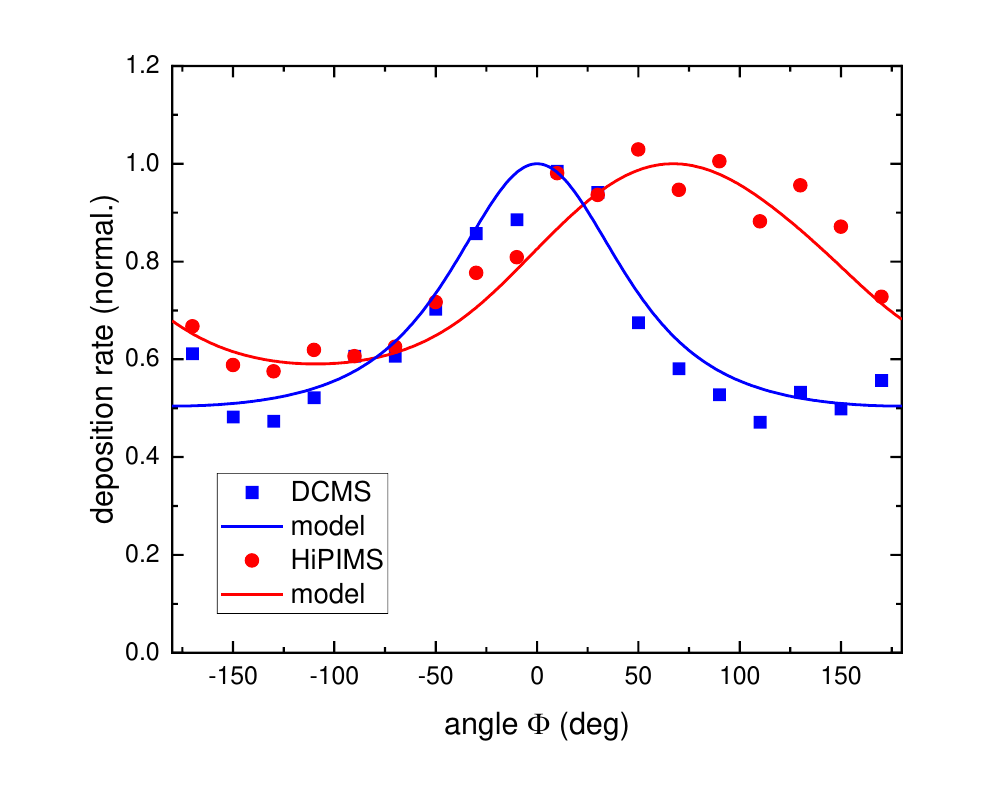}
    \caption{Re-deposition of the marker material along the circumference of the target at different heights $z$ above the target for the DCMS plasma and the HiPIMS plasma.}
    \label{fig:Cr_depositionmodel}
\end{figure}

\begin{itemize}
\item\textit{DCMS:} In the DCMS case, we can assume that the deposition is dominated by neutral species. In a direct-line-of sight model, we assume that the species starting with a cosine distribution at the insert position at ($x_{racetrack}$ = 13.5 mm, $y_{racetrack}$ = 0)  and deposit at the collector surface at a specific height $z$ = 15 mm. The main variation in film growth rate originates from the $1/l^2$ dependence in eq. \ref{eq:depo}. For the azimuthal variation, we have to calculate the different $l$ values for the different angles $\Phi$ from:

\begin{equation}
l(\Phi) = \sqrt{(r_{collector}\cos(\Phi) - y_{racetrack})^2 + \left(r_{collector}\sin(\Phi)\right)^2 + z_{collector}^2}
\end{equation}

and obtain a normalized directed growth rate $\Gamma_{neutrals,direct}$ from neutrals as:

\begin{equation}
\Gamma_{neutrals,direct}(\Phi) = \frac{1}{A}\frac{1}{l(\Phi)^2}
\end{equation}

with $A$ to normalize the distribution $\Gamma_{neutrals,direct}(\phi)$ to the maximum value. In principle, one would also have to regard a  variation in the ejection angle $\theta$ in eq. \ref{eq:depo} along the circumference since the insert is not at the center of the target and the emission angle $\theta$ at $\Phi$ = 0° is larger than the angle $\theta$ at $\Phi$ = 180° for the same height $z$ on the collector. This is, however, an effect of second order because eq. \ref{eq:depo} depends on $\cos(\theta)\cdot\cos(\pi/2-\theta)$, which varies only little for the small range of $\Phi$ where the growth rate is large. By using this scaling, the deposition along the circumference is modelled by a growth rate that is composed of a directed grwoth flux and an un-directed contribution form the background, as shown in Fig. \ref{fig:Cr_depositionmodel} with:

\begin{equation}
\Gamma_{DCMS}(\Phi) = 0.45 + 0.55 \cdot \Gamma_{neutrals,direct}(\Phi)
\end{equation}

Here, we assume that the direct-line-of-sight contributes to 55\% to the deposition rate, whereas 45\% deposition comes from a background chromium deposition. These 45\% may originate from chromium species after scattering with the argon background gas or other chromium species. This leads to a thermalization and to a finite deposition rate everywhere. The agreement between data and this simple model is excellent and highlights that a simple line-of-sight approach is sufficient to describe the data.

\item\textit{HiPIMS:} In the HiPIMS case, we do see a pronounced shift in \ExB direction. Here, we assume that the deposition is governed by ions that experience a significant drag force by the Hall current inducing an azimuthal velocity, as measured above. Such an azimuthal velocity will displace the deposition profile along the circumference in \ExB direction. If we assume for example, that the species start with a typical velocity corresponding to the maximum of the Thompson distribution at an energy of half the surface binding energy $E_{SB}$, we obtain for chromium  ($E_{SB}=1.6 eV$ \cite{husinsky_velocity_1984}) a starting velocity of $v_{sputter}$ = 1.75 km/s. If we then take the time of flight $t$ for such a species until it reaches the collector, we can derive the displacement along the circumference for a given azimuthal velocity $v_{azimuthal}$ as:

\begin{equation}
\Phi_{displacement}(v_{sputter},v_{azimuthal}) = \frac{v_{azimuthal}}{v_{sputter}} \frac{\sqrt{(y_{racetrack}-y_{collector})^2+z_{collector}^2}}{ r_{ractrack}}\label{eq:phishift}
\end{equation}

This yields a displacement of the deposition profile along the circumference of 116°, which is in good agreement with the data. By using eq. \ref{eq:phishift}, we regard the timing of a trajectory only in the $z,y$ plane and the deposition along the circumference is regarded as a rotation of that trajectory by an angle $\Phi_{displacement}(v_{sputter})$. This would neglect that the growth rate at large angles $\Phi_{displacement}(v_{sputter})$ is also affected by the different distances from the insert to the collector. This, however, is taken into account by not rotating individual trajectories, but by rotating the distribution $\Gamma_{neutrals,direct}$, which already contains this distance dependence, as discussed in the following.

The model is also extended by regarding a velocity distribution for the sputtered species and averaging over the different displacements for the different values for $v_{sputter}$ weighted with the amplitude of the Thompson distribution $f_{Thompson} \propto E/(E+E_{SB})^3$ for the ejection velocity $v_{sputter}$ or species energy $E$ and regarding energies up to 100 eV. This yields a normalized growth rate for the ions as:

\begin{equation}
\Gamma_{ions,direct}(\Phi,v_{azimuthal}) = \frac{1}{A}\sum_{E=0}^{E=100 eV} \Gamma_{neutrals,direct}\left(\Phi-\Phi_{displacement}(v_{sputter},v_{azimuthal})\right)f_{Thompson}(E)
\end{equation}

with a normalization $A$, the displacement angle $\Phi_{displacement}(v_{sputter},v_{azimuthal})$ for a given energy of the sputtered species and $v_{sputter}=\sqrt{2E/m}$. For modeling of the actual deposition profile along the circumference,  we have further take into account that ions and neutrals contribute to film growth. The ions experience most significantly the drag force by the Hall current due to their large cross section of ion-electron scattering. Since the ionization degree in HiPIMS is less than 100\% it is conceivable to model the final distribution as an overlap of three components, the background, the directed neutral species, and the azimuthally displaced ions:

\begin{equation}
\Gamma_{HiPIMS}(\Phi) = 0.45 + 0.55 \cdot \frac{1}{A}\left( 0.15\cdot \Gamma_{neutrals,direct}(\Phi) + 0.85 \cdot\Gamma_{ions,direct}(\Phi) \right)
\end{equation}

with $A$ the normalization of the distribution $0.15\cdot \Gamma_{neutrals,direct}(\Phi) + 0.85 \cdot\Gamma_{ions,direct}(\Phi)$. The resulting distribution $\Gamma_{HiPIMS}(\Phi)$ is shown in Fig. \ref{fig:Cr_depositionmodel} with using an azimuthal velocity of  $v_{azimuthal}$ = 1.5 km/s in agreement with the measurement above. It can be seen that the model describe the data very well. Here, we assume that 55\% are deposited in direct line-of-sight and 45\% originates from the background of scattered chromium species. The directed flux consist of deposition from neutrals corresponding to $\Gamma_{neutrals}(\Phi)$ weighted with factor of 0.15 and from the ions $\Gamma_{ions}(\Phi)$ weighted with a factor 0.85, which corresponds to an ionized flux fraction of 85\%, which is typical for such a HiPIMS plasma \cite{biskup_influence_2018}. The small contribution from $\Gamma_{neutrals}(\Phi)$ should also be shifted along the circumference due to the friction between electrons and neutrals in the HiPIMS pulse. The velocity of the neutrals is measured to 0.5 km/s (not shown) and does not vary with distance to the target $z$. This induces a rather small shift that cannot easily be seen in the data.

The match of model and data for the HiPIMS case is excellent, although we regard only a single azimuthal velocity. In principle, one could not only sample the Thompson distribution, but also the measured distribution in $v_{azimuthal}$. However, the agreement is already very good, highlighting that a simple line-of-sight deposition scheme is enough to explain the data. 
\end{itemize}

The good agreement between the simple line-of-sight model and the data is remarkable given the complexity of a HiPIMS plasma. One may argue that the azimuthal movement of the ions is a complex process resulting from the friction of the ions with the electron Hall current. In addition, the transport of ions from the plasma to the collector should be affected by the electric fields in the magnetic pre-sheath or the electric fields in the double layers surrounding the traveling ionization zones, the spokes. Nevertheless, the deposition profile $\Gamma_{HiPIMS}$ still reflects the energetics of the Thompson distribution and the transport time can simply be estimated from the initial velocity of the sputtered species. This is explained in the following:

\begin{itemize}
\item\textit{Azimuthal movement:} The azimuthal velocity of the ions originates from collisions with the electrons in the Hall current along the plasma torus. In these collisions, the momentum transfer is very small and the electrons mainly transfer their kinetic energy to the heavy species and accelerate them in the direction of the \ExB Hall current. Thereby, the initial momentum from the Thompson distribution is preserved and the timing from target to collector is not affected.

The collisions of ions with other heavy species such as argon or metal atoms affect momentum and energy of the ions. This randomizes the trajectories and creates the background contribution to the growth flux that is homogeneously distributed along the circumference. This partial thermalization of the ions by heavy particle collisions is also corroborated by the Maxwellian shape of the velocity distributions \cite{held_velocity_2018,held_velocity_2020}.

\item\textit{Electric fields in the pre-sheath:} Normally, one would expect that the electric fields in the magnetic pre-sheath slow down the ions, so that they return to the target or that their transit time increases leading to a large value for $\Phi_{dsiplacement}$. This may randomize the trajectories and may also contribute to the background deposition along the circumference. Such a background contribution would add to the background from scattering of the ions with other heavy species. Nevertheless, a significant fraction of the deposition flux still reflects the original Thompson distribution of sputtering. This could be understood given the peculiar nature of HiPIMS plasmas: At the chosen power levels, the HiPIMS plasma consists of rotating ionization zones, the spokes, where the ionization rate is maximal, but where the electric field in the magnetic pre-sheath is largely reduced\cite{held_electron_2020}. As a consequence, all sputtered species that are ionized inside a spoke do \textit{not} experience an electric field pointing to the target so that their momentum from the Thompson distribution remains preserved until they reach the collector surface. All sputtered species that are ionized outside of the spoke, do experience the electric fields in the magnetic pre-sheath and return to the target and do, thereby, not contribute to the deposition profile at the collector. 
\end{itemize}

\begin{figure}
    \centering
    \includegraphics[width=0.5\textwidth]{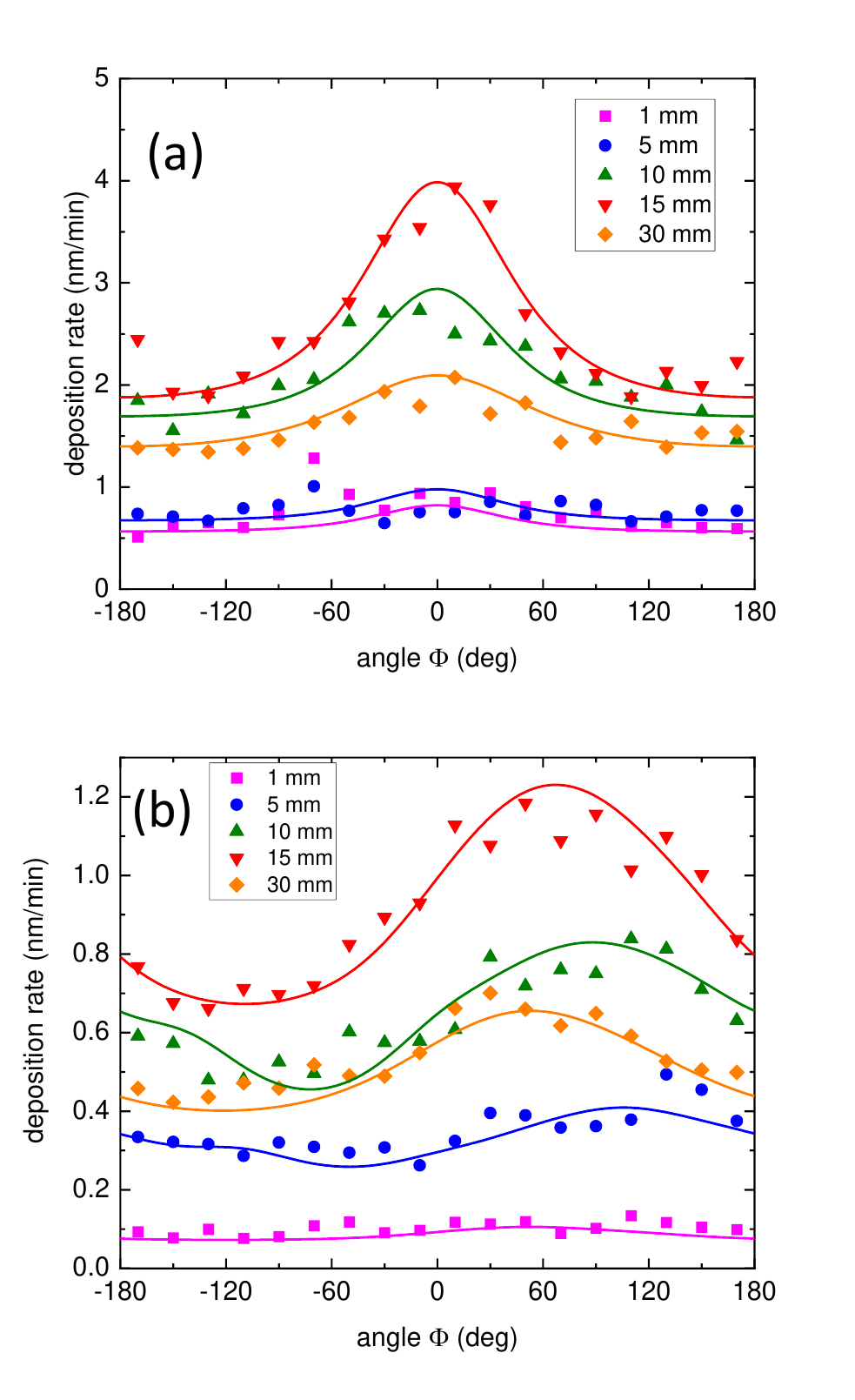}
    \caption{Re-deposition of the marker material along the circumference of the target at different heights $z$ above the target for the DCMS plasma (a) and the HiPIMS plasma (b). The lines denote the model.}
    \label{fig:series}
\end{figure}

This model is now being applied to the measured deposition profiles at the different heights $z$, as shown in Fig. \ref{fig:series}. For this, the directed growth flux in a HiPIMS plasma is again modelled as a combination of the normalized profile $\Gamma_{neutrals}$ to describe 15\%  of the deposition from neutrals and the normalized  profile $\Gamma_{ions}$ to describe 85\% of the deposition from ions. This constitutes the directed growth flux as:

\begin{eqnarray}
\Gamma_{directed,DCMS} &=& \Gamma_{neutrals}\\
\Gamma_{directed,HiPIMS} &=& 0.15 \cdot \Gamma_{neutrals} + 0.85 \cdot \Gamma_{ions}
\end{eqnarray}

\begin{figure}
    \centering
    \includegraphics[width=0.5\textwidth]{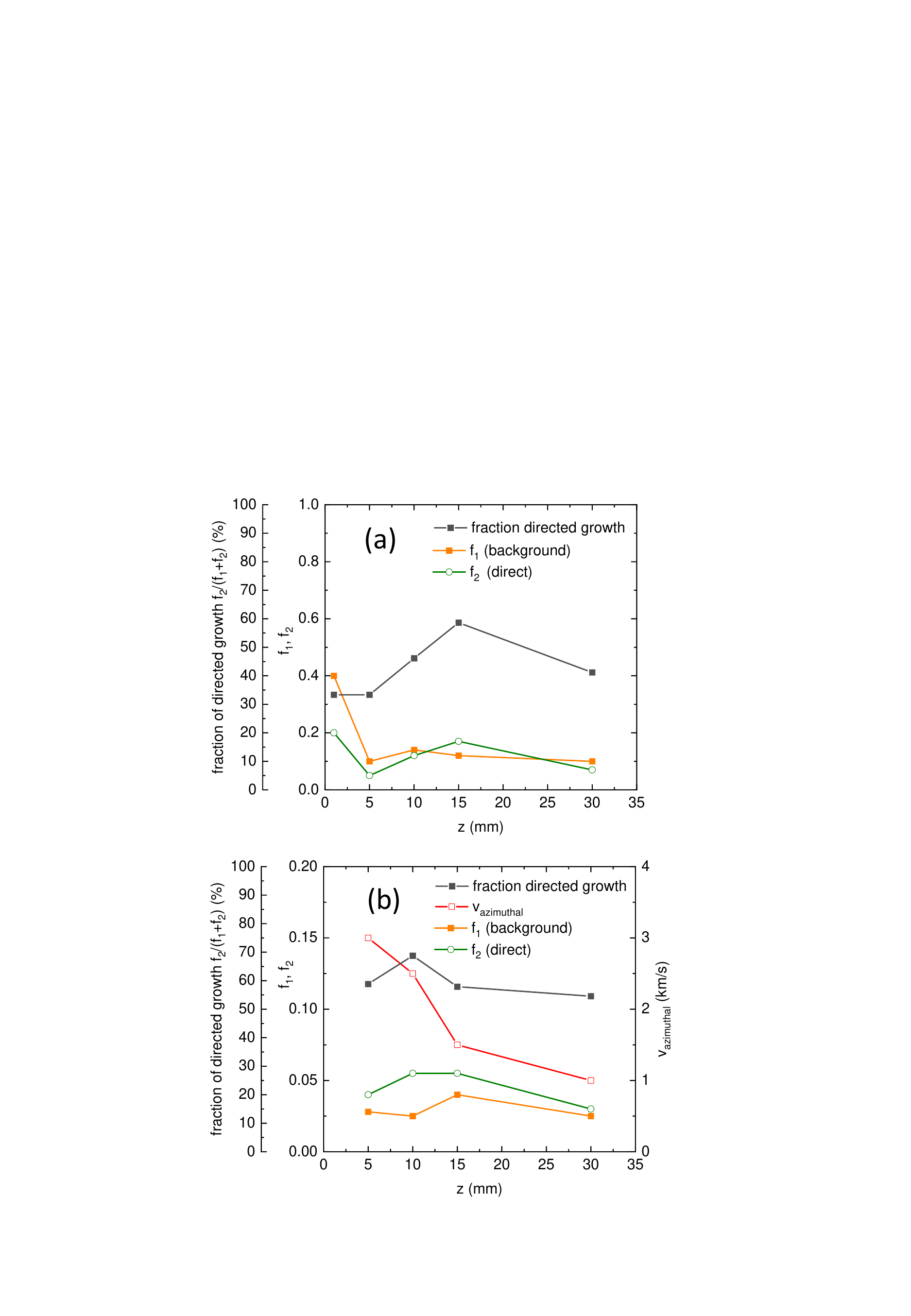}
    \caption{Parameter of the model for the DCMS (a) and HiPIMS (b) deposition profiles showing the scaling factors $f_1$ for the background and $f_2$ for the directed growth flux and the percentage of the directed flux to the growth flux. Azimuthal velocity for modeling the HiPIMS data.}
    \label{fig:seriesparameter}
\end{figure}

Then, the absolute growth rate is composed of the contributions from the background growth flux of scattered metal atoms and ions and the directed line-of-sight growth flux using scaling factors $f_1$ and $f_2$ to yield an absolute growth rate. In addition, the different distances between target and collector at different heights $z$ are corrected by the expression \ref{eq:depo} as:

\begin{equation}
\Gamma_{model} = (f_1 + f_2 \Gamma_{directed})\cos(\theta) \frac{1}{l^2} \cos\left(\frac{\pi}{2}-\theta\right)
\end{equation}

with $l=\sqrt{z^2+(r_{collector}-r_{racetrack})^2}$. Finally, the deposition profiles for the HiPIMS case are adjusted using different azimuthal velocities. This is reasonable, because one can assume that the species experience different friction forces on their trajectories from target and collector (the trajectories are illustrated by the dashed lines in Fig. \ref{fig:sideways_deposition_model}b). The fitting parameters are plotted in Fig. \ref{fig:seriesparameter}.

The modeling of the deposition profiles show very good agreement with a contribution of the directed flux of around 50\% with the largest contribution at a height of $z = \SI{10}{\milli\meter}$ to $z = \SI{15}{\milli\meter}$. Even more interesting is the fact that the HiPIMS data can only be modelled by using different azimuthal velocities. This is already visible in the raw data, with the maximum deposition rate being shifted in \ExB direction to larger angles $\Phi$ for very small values of $z$. For the modelling of the deposition profile at small values of $z$, velocities up to 3 km/s. This could be explained by regarding the different line-of-sight trajectories: At very low values for $z$, the ions pass the longest path within the magnetic pre-sheath above the target. If we regard the electron drift velocity being proportional to $E/B$, a high electric field together with a high electron density in close vicinity to the target will cause a large friction force on the ions, which induce a large displacement $\Phi$. At very large distances $z$, the species spent the smallest time in the magnetic pre-sheath and the accumulation of the friction force is less. It is important to note that the azimuthal velocities in the model are not directly the velocities measured by spectroscopy, because these are measured directly above the racetrack and not on the trajectory from target to collector.

Summarizing, one can state that the simple line-of-sight model can describe the deposition profiles surprisingly well and that the Thompson distribution of the sputtering process is preserved for typically \SI{50}{\percent} of the species reaching the substrate. The very intense HiPIMS plasma only affects their azimuthal displacement.

\section{Conclusion}

By using a chromium insert embedded in a titanium magnetron target, the particle trajectories from the target surface to a collector at the circumference have been tracked. The azimuthal velocities of titanium and chromium ions were obtained from their velocity distribution functions (VDF) measured by optical emission spectroscopy. In the case of titanium ions, the azimuthal velocity is symmetrical distributed around the target in radial direction parallel to the target surface. On the other hand, the chromium ions are accelerated by \SI{0.8}{km\per\second} on their way from the insert position to the other side of the racetrack by the electron Hall current. The chromium ions are slightly faster than the titanium ions since the acceleration is more efficient with chromium than with titanium due to the higher mass of chromium, as can be seen from the different widths of the VDFs.

Measuring the material composition of the coating on the collector around the circumference, the transport of the insert material chromium was analyzed. The transport can be described as a simple line-of-sight process with a starting velocity sampled from the Thompson distribution since all metal species originate from a sputtering process. In the case of a HiPIMS pulse, the strong Hall current causes a uniform azimuthal displacement of the sputtered particles, but their momentum and energy from the sputter process remains preserved. These experiments show that the complexity of a HiPIMS plasma is not significantly affecting the distribution functions of the growth flux. 

\section{Acknowledgement}

This work has been funded by the DFG within the framework of the collaborative research centre SFB-TR 87. %The paper reports on the work being performed in the 3rd funding period of the subproject C7 in the SFB-TR 87.

\section*{Data availability}

The data that support the findings of this study are openly available at the following DOI: 10.5281/zenodo.7904634

\printbibliography

\end{document}